\documentstyle[aps,eqsecnum,12pt]{revtex}
\def\vecq{\vec{\rm\bf q}}
\def\vecqq{\vec{\rm\bf Q}}
\def\Frac#1#2{\frac{\displaystyle #1}{\displaystyle #2}}
\begin{document}
\title{Relativistic three--dimensional two-- and three--body equations \\
on a null plane \\ and applications to meson and baryon Regge trajectories}
\vskip .50in
\author{E. Di Salvo}
\address{Dipartimento di Fisica and I.N.F.N. - Sez. Genova,
Via Dodecaneso,33 - 16146 Genova, Italy}
\author{L. Kondratyuk}
\address{Institute of Theoretical and Experimental Physics,
B.~Cheremushkinskaya ul.~25, 117259 Moscow, Russia}
\author{P. Saracco}
\address{I.N.F.N. - Sez. Genova, Via Dodecaneso,33 - 16146 Genova,
Italy}
\date{\today}
\maketitle
\vskip 1.00in
\begin{abstract}
We start from a field-theoretical model of zero range approximation to
derive three-dimensional relativistic two- and three-body equations on a
null plane. We generalize those equations to finite range interactions.
We propose a three-body null-plane equation whose form is different from the
one presented earlier in the framework of light-cone dynamics. We discuss the
choices of the kernels in two- and three-body cases and
apply our model to the description of meson and baryon Regge trajectories.
Our approach overcomes some theoretical and phenomenological difficulties
met in preceding relativized treatments of the three-body problem.
\end{abstract}
\pacs{xxx}
\vskip .50in
{}~~~~~ \ ~~~~~~ \ ~~~~~~~~ \ ~~~~ \ ~~~~~ \ GEF - Th - 8/1994
\newpage
\baselineskip=1.2\baselineskip
\section{Introduction}
It is very well known that nonrelativistic potential models can be successfully
applied to the description of the heavy $q \overline q$ spectra
(see ref.\cite{ref1} and the recent review \cite{ref2}). However when
the states
containing light quarks (u,d,s) are considered relativistic effects have to be
taken into account ( see, e.g.\cite{ref3,refCaIs2,ref4,ref5}).
Relativistic few body problem
has received a great attention in hadronic and nuclear physics
and a lot of papers have been devoted to the
problem ( see e.g. reviews\cite{ref6,ref7,ref8} and references therein).
Different methods for deriving relativistic few-body three-dimensional
equations have been
discussed. Some authors use a diagrammatic approach, i. e., they
select some leading diagrams and project them onto the
three-dimensional momentum space. Others make use of effective
Hamiltonians, employing various assumptions for their
constructions.

The most appealing are the relativistic approaches based on the
null-plane (or light-cone) dynamics or, equivalently, on the analysis of
Feynman diagrams in the infinite momentum frame (IMF) ( see  recent
papers\cite{ref9,ref9a} and references therein).
One of the main advantages of these  approaches is the very well known fact
that the
wave function of the bound state has a simplest form in IMF, where pair
creation from vacuum is suppressed. This was stressed a long time ago by
Weinberg\cite{ref10}. Indeed in the covariant representation Lorentz invariance
of the S matrix at any order of perturbation theory is ensured by the
contributions containing different numbers of particles and antiparticles in
intermediate states. At the same time diagrams of noncovariant perturbation
theory in IMF are dominated by the simplest intermediate states where all the
particles have positive energies.

To make calculations realistic we must
restrict the number of degrees of freedom. This is usually done by imposing
cutoff on the number of particles and considering only $q\overline q$ sector in
the case of mesons and $3q$ sector in the case of baryons. In order to derive
a two-body null plane equation in the three-dimensional form, one can
use a Hamiltonian approach in which the null planes $\{x_-=\frac{1}{\sqrt{2}}
\ (x^0+x^3)=const\}$ play the role of equal time surfaces
\cite{ref6,ref9,ref9a,ref11}, or eliminate relative null plane
time $x_-$ in the covariant approaches \cite{ref6,ref8,ref12,ref13,ref14}.
However in such approaches it is not trivial to define the angular momentum
operators \cite{ref6,ref8,ref11}. This problem is circumvented in the papers
which use the Relativistic Hamiltonian Dynamics (RHD) in the null plane form
(or Null Plane Dynamics, NPD, or even Light Cone Dynamics, LCD), where
the Poincar\'e generators for the system of two or three interacting particles
are directly constructed in terms of internal variables describing a system
(see \cite{ref6,ref7,ref8,ref11}).

Generally there is no exact correspondence between the results based on RHD or
NPD and quantum field theory (QFT). Such a correspondence can be found in the
Zero Range Approximation (ZRA), where the relative time can be explicitly
eliminated and two-body three-dimensional  null-plane equation can be derived
from the Bethe-Salpeter equation\cite{ref8,ref14}. In this paper we discuss a
generalisation of this equation to finite range interactions, which is
equivalent to a representation of NPD, where the angular momentum operators
have the same form as for free particles\cite{ref11}, so that, under rather
general assumptions, covariance\cite{ref6} of the theory is guaranteed.
Although our results on the two-body null plane equation
are not new, the discussion we present here gives new insight on the relation
between NPD and QFT. Moreover we relate the kernel in the
relativistic two-body null plane equation to the potential used by Godfrey and
Isgur\cite{ref3}; this, together with linearity of meson Regge trajectories,
suggests that for large separations between constituents the interaction should
be of the oscillator type, whose parameter is proportional to the string
tension squared. Such a kernel leads to predictions on Regge
trajectories for mesons composed of light ($u$, $d$ and $s$) quarks which are
in good agreement with data, including small deviations from linearity for
strange mesons.

Another new point in our paper is the derivation of three-body null plane
equations. Starting from the covariant three-body equation, we
show that in ZRA two relative times can be eliminated and a three-dimensional
three-body null plane equation  for bound or scattering cases
can be derived.
We consider also  a simple way of generalizing our treatment to finite
range interactions, in line with Relativistic Hamiltonian Dynamics.

The form of the three-body null plane  equations  is different from
the one proposed earlier in the framework of RHD\cite{ref15},
as well as from the naive relativistic generalization of the Schr\"odinger
equation used in refs.\cite{ref3,refCaIs2,refCaIs} for the description of
baryon spectra. In particular, as we show in the present paper, our
approach presents theoretical and also phenomenological advantages over these
treatments.

We apply  three-body null plane  equations to Regge trajectories for a system
of three relativistic quarks, analyzing the kernel corresponding
to a three-body force with string junction, involved in the mass operator
squared. We take this operator in a form which commutes with angular momentum
in a representation which coincides with the one for free particles, so that
also the three body equation yields a covariant\cite{ref6}  description of
dynamics. Approximate linearity
of Regge trajectories fixes the dependence of the kernel on
two different relative coordinates, which for large relative separations has to
be still of the oscillator type. Equating  the slopes of two Regge
trajectories corresponding to the orbital excitations of two relative
degrees of freedom fixes the sector that describes the relativistic recoil of
the moving two-quark subsystem. The implementation of the condition that
the slope of the diquark-quark orbital excitations should be same as for
meson Regge trajectories fixes the oscillator parameter. We apply the model to
the baryon Regge trajectories composed from the light quarks and show that the
predictions of our model are in good agreement with data, reproducing small
deviations from linearity.

It is worth stressing that, unlike the
approaches based on a relativized Schr\"odinger equation, our three-body
Hamiltonian satisfies cluster separability for two-body forces;
furthermore,
implementing our model with three-body interactions allows to pass from mesons
to baryons, without re-adjusting fundamental parameters or assuming a different
kind of interaction between quarks.

The paper is organized as follows.  In Sect.~2
we carry on the program of studying
the relativistic two-body equation on a null plane. Starting from
the covariant Bethe-Salpeter equation and from a field theory model in ZRA,
we derive a relativistic two-body equation for a null-plane wave-function.
Then we generalize this equation for finite range
interactions and introduce the relativistic null-plane Lippmann-
Schwinger equation.

In Sect.~3 we focus on the three-body equation. As in the two-body case,
we start from ZRA; then, using the form of the equation
introduced for finite range interaction in the two-body case, we
formulate a relativistic equation for a three-body system
appropriate for this kind of interactions. Moreover we show
that this equation satisfies cluster separability condition.

In Sects.~4 and 5 we apply two- and three-body equations
to the study of meson and
baryon Regge trajectories. We  discuss the choice of the
kernels in two- and three-body cases and we apply the model to the
description of the experimental data of meson Regge trajectories composed of
$q\overline q$, $q\overline s$ (or $s\overline q$) and $s\overline s$
orbital excitations and Regge trajectories of baryons composed of
$u$- and/or $d$-quarks. We consider also the $\Lambda$ Regge trajectory.

Sect. 6 is devoted to conclusions.

\section{Two - body null plane equation}

Let us consider the case of two non-identical, spinless interacting particles,
with masses $m_1$ and $m_2$ respectively. The Bethe-Salpeter (BS) equation for
the reduced amplitude $\chi$ reads\cite{refBeSa,refItZu}
\begin{equation}
 (p_1^2 - m_1^2) (p_2^2 - m_2^2) \chi (p_1, p_2) +
\int \frac{d^4p_1'}{(2\pi)^4}
 \frac{d^4p_2'}{(2\pi)^4} (2\pi)^4 \delta^{4} (P - p_1' - p_2') U_{12}
\ (p, p'; P) \chi(p_1', p_2') = 0 , \label{due-uno}
\end{equation}
where
\begin{equation}
 p = \eta p_1 - (1-\eta) p_2, \ ~~~~ \  p' = \eta p_1' - (1-\eta) p_2'
\ ~~~~~ \ (0 < \eta < 1)
\end{equation}
and $U_{12}$ is the kernel in momentum space.

The BS equation meets some intrinsic difficulties.
First of all,  the normalization condition for $\chi$
\begin{equation}
\int \frac{d^4p}{(2\pi)^4} \frac{d^4p'}{(2\pi)^4}
{\overline \chi} (p_1,p_2){d \over dP_{\mu}} [(p_1^2 - m_1^2) (p_2^2 -
m_2^2) (2\pi)^4
\delta^4 (p'-p) + U_{12} (p, p'; P)] \chi (p_1,p_2) = 2P_{\mu}
\label{due-due}
\end{equation}
depends on the interaction through the kernel $U_{12}(p,p';P)$, which is
a source of complications.

Moreover $\chi$ depends on the relative energy, therefore $\chi_P (x)$
depends on the relative time $t$. This property of the reduced
amplitude $\chi$ leads to different complications when we try to
apply the BS equation to physical problems. Indeed, in the ladder
approximation the BS approach does not exhibit the correct static
limit, i. e., when, e. g., $m_1$ becomes large, while keeping $m_2$
fixed: some solutions exist with negative norm or which do not
satisfy the correct non-relativistic limit. This problem can be
avoided by eliminating the relative time dependence\cite{ref8,ref14}.

Let us define a null plane projection of the BS amplitude as follows:
\begin{equation}
\psi (  {\rm {\bf p}}_1,   {\rm {\bf p}}_2) = \int dp_{1-} dp_{2-}
p_{1+} p_{2+} \delta (P_{-}-p_{1-}-p_{2-}) \chi (p_1, p_2) \ ,
\label{due-tre}
\end{equation}
where ${\rm {\bf p}}_i$ are null plane three-momenta, i. e.,
\begin{equation}
{\rm {\bf p}}_i \equiv ({\rm {\bf p}}_{iT}, p_{i+})
\label{due-imp}
\end{equation}
and
\begin{equation}
p_{i+} = {{E_i+p_{iz}} \over {\sqrt{2}}}, \ ~ \ {\rm {\bf p}}_{iT} \equiv
(p_{ix}, p_{iy}) \ . \nonumber
\end{equation}

We assume the kernel of eq. (\ref{due-uno}) to be independent
of relative null-plane energy $p_-$, i. e.,
\begin{equation}
U_{12} = U_{12} (  {\rm {\bf p}}, {\rm {\bf p}}'; P) , \label{due-quattro}
\end{equation}
where
\begin{equation}
{\rm {\bf p}} = \eta {\rm {\bf p}}_1 - (1-\eta){\rm {\bf p}}_2 , \ ~~~~~~ \
{\rm {\bf p}'} = \eta {\rm {\bf p}'}_1 - (1-\eta){\rm {\bf p}'}_2 .
\nonumber
\end{equation}
This amounts to assuming an instantaneous interaction in the null plane time
$x_+$. Then the reduced BS amplitude may be rewritten as
\begin{equation}
\chi (p_1,p_2) = -{1 \over \pi}{{\int d \Gamma_{12}
U_{12} ({\rm {\bf p}}, {\rm {\bf p}'}; P)
\psi ({\rm {\bf p}'}_1,{\rm {\bf p}'}_2)} \over {(p_1^2 -
m_1^2) (p_2^2 - m_2^2)}} ,
\label{due-cinque}
\end{equation}
where
\begin{equation}
d \Gamma_{12} = {1 \over {2(2\pi)^3}}{{d {\rm {\bf p}}'_1} \over p'_{1+}}
{{d {\rm {\bf p}}'_2} \over p'_{2+}} \delta^3 ({\rm {\bf P}} -
{\rm {\bf p}'}_1 - {\rm {\bf p}'}_2) .
\nonumber
\end{equation}

Integrating both sides of (\ref{due-cinque}) over
$p_{1+}p_{2+}dp_{1-}dp_{2-} \delta (P_{-}-p_{1-}-p_{2-})$, we
derive the following equation for a null plane wave-function:

\begin{equation}
\psi ({\rm {\bf p}}_1,{\rm {\bf p}}_2) =
i P_+ G_0 (x_1,{\rm {\bf p}}_{1T};x_2,{\rm {\bf p}}_{2T})
\int d \Gamma_{12} U_{12} ({\rm {\bf p}},{\rm {\bf p}'};P) \psi ({\rm
{\bf p}'}_1,
{\rm {\bf p}'}_2) ,
\label{due-sei}
\end{equation}
where
\begin{equation}
G_0 (x_1,{\rm {\bf p}}_{1T}; x_2,{\rm {\bf p}}_{2T})
= - [{{\rm {\bf p}_{1T}^2+m_1^2} \over x_1}+{{\rm {\bf
p}_{2T}^2+m_2^2} \over
x_2} - M^2 - {\rm {\bf P}}_T^2 - i \epsilon]^{-1}
\label{due-sette}
\end{equation}
and
\begin{equation}
M^2 = P^2, \qquad x_i = {p_{i+} \over P_+} ,
\qquad
{\rm {\bf P}}_T =
{\rm {\bf p}}_{1T} + {\rm {\bf p}}_{2T}.
\label{due-otto}
\end{equation}
Moreover let us assume the kernel $U_{12}$ to be of the type
\begin{equation}
U_{12} = U_{12} (x, {\rm {\bf p}}_T; x', {\rm {\bf p}'}_T) \ ,
\label{pot-type}
\end{equation}
where
\begin{equation}
\qquad x = \eta x_1 - (1-\eta)x_2, \ ~ \ ~ \
{\rm {\bf p}}_T = \eta {\rm {\bf p}}_{1T} - (1-\eta){\rm {\bf p}}_{2T} \ .
\label{variabili}
\end{equation}
Then we may define a null-plane wave-function $\Psi$ that, according to
(\ref{due-sei}), satisfies the following equation:
\begin{eqnarray}
[{{{\rm {\bf p}}_{1T}^2 + m_1^2} \over x_1}+{{{\rm {\bf
p}}_{2T}^2 + m_2^2}
\over x_2} - M^2 - {\rm {\bf P}}_T^2] \Psi (x_1,{\rm {\bf
p}}_{1T} ;
x_2,{\rm {\bf p}}_{2T}) + \nonumber\\
 i  \int d L'_{12} U_{12} (x,{\rm {\bf p}}_T ; x',{\rm {\bf p}'}_T)
\Psi
(x'_1,{\rm {\bf p}'}_{1T} ; x'_2,{\rm {\bf p}'}_{2T}) = 0 ,
\label{due-nove}
\end{eqnarray}
where
\begin{equation}
d L'_{12} = {1 \over {2(2\pi)^3}}[\prod_{r=1}^2 {{dx'_r} \over {x'_r}} d{\rm
{\bf p}'}_{rT}]  \delta (1-x'_1-x'_2) \delta^2
({\rm {\bf P}}_T-{\rm {\bf p}'}_{1T}-{\rm {\bf p}'}_{2T}) .
\label{due-dieci}
\end{equation}

At this point we discuss approximation
(\ref{due-quattro}). The simplest case is ZRA, i. e., when $U_{12}$ is a
constant. In this case the null plane wave-function reads
\begin{equation}
\Psi  = \frac{\displaystyle N}
{
\frac{\displaystyle{\rm\bf q}_T^2+m_1^2}{\displaystyle \xi}
+\frac{\displaystyle{\rm\bf q}_T^2+m_2^2}{\displaystyle 1-\xi}
-M^2-i\varepsilon
},
\end{equation}
where we have introduced the variables
\begin{equation}
\qquad \xi = x_1,  \ ~ \ ~ \
{\rm {\bf q}}_T = (1-\xi) {\rm {\bf p}}_{1T} - \xi {\rm {\bf p}}_{2T} \,
\label{due-ottoprimo}
\end{equation}
and the constant $N$ is determined by the normalization condition
{}~(\ref{due-due}), i. e.,
\begin{equation}
\frac{1}{2(2\pi)^3}\int {{d \xi d {\rm {\bf q}}_T} \over {\xi (1-
\xi)}} |\Psi (\xi,
{\rm {\bf q}}_T)|^2 = 1 ,
\nonumber
\end{equation}
which turns out to coincide with the usual normalization of the
infinite momentum frame wave-function.

In this limit a null plane description is equivalent to field theory;
however it has limited applications. It is therefore important to
construct kernels of finite size, using different models. For example, it is
possible to use ladder approximation to the BS equation and to project it on
the null plane. Such an equation was suggested by Weinberg \cite{ref10}.
Another example was considered by t'Hooft
\cite{ref12}, who derived a two-body null plane equation for the $q \overline
q$ system in $1+1$ QCD, taking the limit of $N_c \rightarrow \infty$, in such a
way that $\alpha_s N_c$ have a finite, constant value.

A great deal of papers were devoted to the analysis of different aspects of
Hamiltonian null plane approach \cite{ref9}. Here we try to impose the
Tamm-Dancoff cutoff on the number of particles, while conserving a very simple
representation of angular momentum operators. Generally speaking, in null plane
formalism angular momentum operators should depend on interaction
\cite{ref6,ref8};
in particular this is the case of the Weinberg equation \cite{ref10}, where the
representation of angular momentum operators cannot be defined
straightforwardly, since these operators would depend on interaction in a very
complicated way. Furthermore, as stressed by Leutwyler and
Stern\cite{ref6}, without defining angular momentum, the wave function cannot
be normalized in a usual way. Therefore the normalization condition for the
wave-function satisfying the Weinberg equation is not clear, except in ZRA,
where the S-wave alone is involved.

However, as shown by Terent'ev \cite{ref11}, in the two-body sector with a
finite range of interaction it is also possible to choose for the angular
momentum operators the same representation as for free particles. For example,
in the relativistic oscillator model\cite{refFey} the two
different representations are related by a "gauge" transformation \cite{ref8}.
Usually one assumes this equivalence to hold true also for other interactions.
In the Terent'ev representation the angular momentum operator takes the form
\begin{equation} \vec{\rm\bf l} = -i \vec{\rm\bf q}  \times
\ \frac{\partial}{\partial\vec{\rm\bf q}} ,
\label{angmom}
\end{equation}
where
\begin{eqnarray}
\vecq  &\equiv& \left({\rm\bf q}_T,q_z\right),\qquad
q_z = (\xi - {1 \over 2}) M_{12}  -
{{m_1^2-m_2^2} \over {2 M_{12}}},
\label{vect}
\\
M_{12} &=& M_{12}(\vecq) = \epsilon_1+\epsilon_2, \label{dopoancora}
\\
\epsilon_i &=& \sqrt{\vec{\rm\bf q}^2 + m_i^2} , \ ~~ \ i=1,2
\end{eqnarray}
and the variables $\xi$ and $\rm {\bf q}_T$ are defined by
eqs.~(\ref{due-ottoprimo}).
In this representation angular momentum conservation is ensured by rotation
invariance of the kernel in eq. (\ref{due-nove}).

Then it is possible to rewrite eq. ~(\ref{due-nove}) in the form
\begin{equation}\hat M^2_{12} \Psi = M^2 \Psi \ ,
\label{due-dodici}
\end{equation}
where
\begin{equation}
\hat M^2_{12} = M_{12}^2(\vecq) + W_{12}, \label{ciservedopo}
\end{equation}
and
\begin{equation}
W_{12} = i U_{12}.
\nonumber\end{equation}
If $W_{12}$ commutes with the
angular momentum operator, the angular condition is satisfied and a
covariant\cite{ref6} description of the composite system is guaranteed.

Extension to the case or two interacting spinning particles can be performed
in a straightforward way by using the Terent'ev representation \cite{ref11}.
In this case the meson state vector can be written in the form
\begin{eqnarray}
\mid P_+, {\rm\bf P}_T, J, \lambda >_{NP} =
\nonumber\\
\sum_{ln\sigma} \sum_{\nu' \mu' \nu \mu} C^{1 \lambda}_{ln,1\sigma}
C^{1 \sigma}_{{1 \over 2} \mu,{1 \over 2}\nu}
\ \int {{d^2 q_T} \over P_+} \frac{d\xi}{\xi} \phi_l (q) Y_{ln} (\hat q)
\ \hat V_{\mu'\mu}
\ (\vec{\rm\bf q}) \hat V_{\nu'\nu} (-\vec{\rm\bf q}) |p_1 \mu', P-p_1
\nu'>_{NP},
\nonumber
\end{eqnarray}
where NP means "null plane", the $C$'s are Clebsch-Gordan coefficients,
$Y_{ln}$ the Legendre functions,
$\vec{\rm\bf q}$ is defined by eq. (\ref{vect}), $p_1 = p_1(\vec{\rm\bf q})$ is
the four-momentum of the quark,

\begin{equation} q = \mid \vec{\rm\bf q} \mid, \ ~~~ \
\ \hat q = \frac{\vec{\rm\bf q}}{q}
\end{equation}

and the $\hat V$'s are the Melosh\cite{ref17} rotation matrices for quark and
antiquark, i. e.\cite{ref11},

\begin{equation}
\hat V (\vec{\rm\bf q}) = \frac{\displaystyle
\ m_1+\epsilon_1+q_z+i\epsilon_{js}\sigma_j q_s}{\displaystyle \left[ 2
(\epsilon_1+m_1)(\epsilon_1+q_z) \right]^{1 \over 2}}
\end{equation}

\begin{equation}
\hat V (-\vec{\rm\bf q}) = \frac{\displaystyle
m_2+\epsilon_2-q_z-i\epsilon_{js}\sigma_j q_s}{\displaystyle \left[ 2
(\epsilon_2+m_2)(\epsilon_2-q_z) \right]^{1 \over 2}}
\end{equation}

The state vector $\mid P_+, {\rm\bf P}_T, J, \lambda >_{NP}$
at four-momentum $P$
is obtained from the state vector at rest
$\mid P_{0+}, {\rm\bf 0}_T, J, \lambda >_{NP}$ by the element of the
Poincar\'e
group $l (P \leftarrow P_0)$ equivalent to the product of two Lorentz
transformations, i. e., $\lambda (P \leftarrow P_{\infty})
\ \lambda(P_{\infty} \leftarrow P_0)$, where $P_0 \equiv (M,0,0,0)$.

Now let us consider nucleons interacting through exchange of different mesons.
The kernel that describes such an interaction can be identified
with the Born term expressed in terms of null - plane variables.
For example, in the cases of scalar and vector meson
exchange, the kernels have, respectively, the following forms\cite{refKoTe}:
\begin{equation}
W_{12s,v} = {\cal V}^+ (\vec{\rm\bf q'}) W_{s,v}^{(r)}(\vec{\rm\bf q'},
\ \vec{\rm\bf q}) {\cal V}(\vec{\rm\bf q}),
\end{equation}
where
\begin{equation}
{\cal V} (\vec{\rm\bf q}) =\hat V (p_1) \otimes \hat V (p_2) \, , ~~ \ ~ \ ~ \
{\cal V} (\vec{\rm\bf q}') =\hat V (p'_1) \otimes \hat V (p'_2)\, ,
\end{equation}
\begin{equation}
W_{12s}^{(r)} (\vec{\rm\bf q'},\vec{\rm\bf q}) =
\ \frac{\displaystyle g^2 }{\displaystyle (\vec{\rm\bf q'} -
\ \vec{\rm\bf q})^2+\mu^2} {\overline u} (p'_1) u(p_1)
\ {\overline u} (p'_2) u(p_2), \ ~ \ p^{(')}_i \equiv (\epsilon_i^{(')},
\ (-)^{1+i} \vec{\rm\bf q}^{(')}),
\end{equation}
\begin{equation}
(W_{12v}^{(r)})_{\alpha'\beta',\alpha\beta}(\vec{\rm\bf q'},\vec{\rm\bf q})
\ =  \frac{\displaystyle g^2 }{\displaystyle (\vec{\rm\bf q'} -
\ \vec{\rm\bf q})^2+\mu^2}
\  V^{\mu}_{\alpha\beta}(\vec{\rm\bf q'},
\ \vec{\rm\bf q}) V^{\nu}_{\alpha\beta}(-\vec{\rm\bf q'}, -\vec{\rm\bf q})
\ g_{\mu\nu},
\nonumber
\end{equation}

\begin{equation}
V^i_{\alpha\beta}(\vec{\rm\bf q'}, \vec{\rm\bf q}) = (\vec{\rm\bf q'}+
\ \vec{\rm\bf q}) \delta_{\alpha\beta}+i\left[(\vec{\rm\bf q'}-\vec{\rm\bf q})
\ \times \vec{\bf \sigma} \right]_{\alpha\beta} \, , \ ~~ \ (i= 1,2),
\end{equation}

\begin{equation}
V^0_{\alpha\beta}(\vec{\rm\bf q'}, \vec{\rm\bf q}) = \frac{\displaystyle
\ \left[ (\epsilon+m)^2 + \vec{\rm\bf q} \cdot \vec{\rm\bf q'} \right]
\ \delta_{\alpha\beta}+i\vec{\bf \sigma}_{\alpha\beta} \cdot \vec{\rm\bf q}
\ \times \vec{\rm\bf q'}}{\displaystyle \epsilon+m} \, ,
\end{equation}

$\mu$ is the mass of the exchanged meson and $\vec{\rm\bf q'}$ is defined
analogously to $\vec{\rm\bf q}$ (eq. \ref{vect}), with $M_{12}$
substituted by $M'_{12}$, which generally assumes a different value. In
the case of
vector meson exchange we have assumed the two interacting particles to have
equal masses, $m_1 = m_2 =m$, therefore we have also $\epsilon_1 = \epsilon_2
= \epsilon$. In the Terent'ev representation of angular momentum, such kernels
satisfy the angular condition\cite{ref15}.

If we consider kernels of this kind, there is no exact
correspondence between null plane dynamics in two - body sector
and field theory; however the former description may be a good
approximation to the latter one if only two -
body intermediate states are important. A further improvement
would consist in constructing multichannel null - plane dynamics;
indeed, if the number of channels becomes infinite, null - plane
dynamics and field theory will turn out to be equivalent.

Equation (\ref{due-nove}) may be also written symbolically as
\begin{equation}
\Psi = G_0 W_{12} \Psi,
\label{due-tredici}
\end{equation}
where $G_0$ is defined by eq. (\ref{due-sette}).
This leads immediately to the Lippmann-Schwinger equation for
the two body relativistic T - matrix, i. e.,
\begin{equation}
T = W_{12} + W_{12} G_0 T,
\nonumber\end{equation}
which, in momentum representation, reads
\begin{eqnarray}
T (x_1,{\rm {\bf  p}}_{1T},x_2, {\rm {\bf  p}}_{2T}; x'_1,
{\rm {\bf  p}'}_{1T}, x'_2, {\rm {\bf  p}'}_{2T}) =
W_{12} (x,{\rm {\bf  p}}_T; x', {\rm {\bf  p}'}_T) +
\nonumber\\
\ ~~~~~~~~~~~~~~~~
\nonumber\\
+ \int d L''_{12}
{{W_{12} (x,{\rm {\bf  p}}_T; x'',{\rm {\bf  p}''}_T) T (x''_1,{\rm {\bf
p}''}_{1T}, x''_2,{\rm {\bf  p}''}_2; x'_1, {\rm {\bf  p}'}_{1T},  x'_2,
{\rm {\bf  p}'}_{2T})} \over
{M^2+{\rm {\bf P}}_T^2-{{{\rm {\bf p}''}_{1T}^2 + m_1^2} \over
x''_1}-
{{{\rm {\bf p}''}_{2T}^2 + m_2^2} \over x''_2} + i \epsilon}} .
\label{due-quattordici}
\end{eqnarray}

As proved in ref. \cite{ref11}, in this scheme angular
momentum operators are the same as for free particles, the only
difference consisting in the appearence of Melosh rotation
matrices in partial wave expansion of the
wave-function and of the scattering amplitude.

\section{Three-body null plane equation}
Analogously to the two-body case, the reduced three-body amplitude reads

\begin{eqnarray}
(p_1^2 - m_1^2) (p_2^2 - m_2^2) (p_3^2 - m_3^2) \phi (p_1,
p_2, p_3)+
\nonumber\\
\ ~~~~~~~~
\nonumber\\
+ \int \frac{d^4p'_1}{(2\pi)^4}
\frac{d^4p'_2}{(2\pi)^4} \frac{d^4p'_3}{(2\pi)^4}
(2\pi)^4 \delta^{4} (P - p'_1 - p'_2 -
p'_3)U_{123} (\Pi, \Pi'; P) \phi (p'_1, p'_2, p'_3)  = 0,
\label{tre-uno}
\end{eqnarray}
where $U_{123}$ is the kernel and
\begin{equation}
\Pi \equiv (p_1, p_2, p_3), \qquad
\Pi' \equiv (p_1', p_2', p_3') .
\label{tre-due}
\end{equation}
The null plane wave-function is defined by
\begin{equation} \psi' ({\rm {\bf p}}_1, {\rm {\bf p}}_2, {\rm {\bf
p}}_3) = \int dp_{1-} dp_{2-} dp_{3-}
p_{1+} p_{2+} p_{3+} \delta (P_{-}-p_{1-}-p_{2-}-p_{3-}) \phi (p_1,
p_2, p_3) \ .
\label{tre-tre}
\end{equation}
Now let us suppose the kernel to be independent of $p_{r-}$, i. e.,
\begin{equation} U_{123} = U_{123} ({\rm {\bf \Pi}}, {\rm {\bf \Pi'}}; P) ,
\label{tre-quattro}
\end{equation}
where
\begin{equation}
{\rm {\bf \Pi}}
\equiv ({\rm {\bf p}}_1, {\rm {\bf p}}_2, {\rm {\bf p}_3}) ,
\quad
{\rm {\bf \Pi'}} \equiv
({\rm {\bf p}'}_1,{\rm {\bf p}'}_2, {\rm {\bf p}'}_3) .
\label{tre-cinque}
\end{equation}

In this case  we derive from (\ref{tre-uno}) the following equation
for null plane three-body wave-function:
\begin{equation}\psi' ({\rm {\bf \Pi}})  =
- G'_0 (x_i, {\rm {\bf p}}_{iT})
\int d L'_{123} U_{123} ({\rm {\bf \Pi}},{\rm {\bf \Pi}'};P)
\psi' ({\rm {\bf \Pi}'})
\label{tre-sei}
\end{equation}
 where
\begin{equation}d L'_{123} = {1 \over {4(2\pi)^6}}\prod_{r=1}^3 {{dx'_r} \over
{x'_r}} d{\rm {\bf p}'}_{rT}
\delta (1-x'_1-x'_2-x'_3) {\delta}^2 ({\rm {\bf P}}_T-{\rm {\bf
p}'}_{1T}-
{\rm {\bf p}'}_{2T}-{\rm {\bf p}'}_{3T})
\label{tre-sette}
\end{equation}
and
\begin{equation}G'_0 (x_i, {\rm {\bf p}}_{iT}) =
\left[
\frac{\displaystyle{{\rm {\bf p}}_{1T}^2+m_1^2} }{\displaystyle
x_1}+{{{\rm {\bf
p}}_{2T}^2+m_2^2}
\over x_2} + {{{\rm {\bf p}}_{2T}^2+m_2^2} \over x_2} - M^2 -
{\rm {\bf P}}_T^2 - i \epsilon\right]^{-1} \ .
\label{tre-otto}
\end{equation}

If we assume the kernel $U_{123}$ to depend on $p_{r+}$ only
through fractional momenta $x_r$, the three-body equation reads
\begin{eqnarray}
&\left[
\frac{\displaystyle {\rm \bf p}_{1T}^2+m_1^2 }
{\displaystyle x_1}+
\frac{\displaystyle{\rm \bf p}_{2T}^2+m_2^2}{\displaystyle x_2}
+\frac{\displaystyle {\rm \bf p}_{3T}^2+m_3^2}{\displaystyle x_3}
- M^2 - {\rm {\bf P}}_T^2
\right] \Psi' (x_i, {\rm {\bf  p}}_{iT}) + \nonumber \\ &
+\int d L'_{123}
U_{123} (x_r,{\rm {\bf p}}_{rT};  x'_{s},{\rm {\bf p}'}_{sT})
\Psi' (x'_i, {\rm {\bf  p}'}_{iT}) = 0 \ .
\label{tre-nove}
\end{eqnarray}
We write, as in the two-body case, the Lippmann-Schwinger
equation for three body T-matrix, i. e.,
\begin{eqnarray}
T (x_i, {\rm {\bf  p}}_{iT};  x'_i, {\rm {\bf
p}'}_{iT}) =
U_{123} (x_r, {\rm {\bf  p}}_{rT};  x'_s, {\rm {\bf
p}'}_{sT})+
\nonumber\\
\ ~~~~~~~~~~~
\nonumber\\
+ \int d L''_{123}
\frac{\displaystyle {U_{123} (x_r, {\rm {\bf  p}}_{rT};  x'_{s}, {\rm {\bf
p}'}_{sT})
T (x''_i, {\rm {\bf  p}''}_{iT};  x'_i, {\rm {\bf  p}'}_{iT})
}
}{\displaystyle { \displaystyle
M^2+{\rm {\bf P}}_T^2 -{{{\rm {\bf p}''}_{1T}^{2}+m_1^2} \over
x''_1}-
{{{\rm {\bf p}''}_{2T}^2+m_2^2} \over x''_2} -
{{{\rm {\bf p}''}_{3T}^2+m_3^2} \over x''_3}+ i \epsilon}} .
\label{tre-dieci}
\end{eqnarray}

Now we assume only two-body interactions to occur. Then the
kernel of eq.~(\ref{tre-nove}) may be written as
\begin{equation}
U_{123} (x_{r}, {\rm {\bf  p}}_{rT};  x'_{s}, {\rm {\bf  p}'}_{sT}) =
i \sum_{i \ne j \ne k} x_k U_{ij} ({x_{ij},{\rm\bf p}}_{ijT};
{x'_{ij},{\rm\bf p}'}_{ijT}) 2(2\pi)^3
\delta (x_k-x'_k) \delta^2 ({\rm {\bf p}}_{kT}-{\rm {\bf
p}'}_{kT}),
\label{tre-undici}
\end{equation}
where ${\rm\bf p}_{ijT}$ and $x_{ij}$ are defined analogously to eq.
{}~(\ref{variabili}).
Then equation (\ref{tre-nove}) can be written as
\begin{equation} \hat M^2_{123} \Psi' = M^2 \Psi',
\nonumber
\end{equation}
where we have set
\begin{eqnarray}
\hat M^2_{123}  &=& \sum_{i=1}^3 {{m_i^2+{\rm
{\bf p}}_{iT}^2} \over x_i}+
\sum_{i \ne j \ne k} {W_{ij} \over {1-x_k}} ,
\label{tre-dodici}
\\
W_{ij} \Psi' &=& i \int d L_{ij}^{k'}
U_{ij} (x_{ij},{\rm {\bf p}}_{ijT};  x'_{ij},{\rm {\bf p}'}_{ijT}) \Psi'
({\rm x'_i,{\bf p}'}_{iT};
x'_j,{\rm {\bf p}'}_{jT}; x_k,{\rm {\bf p}}_{kT})
\label{tre-dodiciprimo}
\end{eqnarray}
and $d L_{ij}^{k'}$ is defined by (\ref{due-dieci}), substituting
the overall transverse
momentum ${\rm {\bf P}}_T$ by
\begin{equation}{\rm {\bf P}}_{ijT} = {\rm {\bf p}}_{iT}+{\rm {\bf
p}}_{jT}.
\nonumber\end{equation}

In the case of ZRA $W_{ij}$ are constants and condition (\ref{tre-quattro}) is
satisfied automatically. Such an equation is equally satisfied if the
two--body interactions are of the type discussed in the previous
section. Here we show
that eq. (\ref{tre-nove}), with a kernel like (\ref{tre-undici}),
satisfies cluster separability
condition. Indeed, if, e. g., the third particle is taken far away from
the other two, we have $W_{23} = W_{31} = 0$ and
eq.~(\ref{tre-nove}) reduces to
\begin{equation}
({{{\rm {\bf P}}_{12T}^2+\hat M_{12}^2} \over {1-x_3}} +
{{{\rm {\bf p}}_{3T}^2+m_3^2} \over x_3} - {\rm {\bf P}}_T^2)
\Psi' = M^2 \Psi' ,
\label{tre-tredici}
\end{equation}
where $\hat M_{12}^2$ is given by eq. (\ref{ciservedopo}).

Dividing both sides of eq. (\ref{tre-tredici}) by $2P_+$, we get
\begin{equation}
P_- \Psi' = (P_{12-} + p_{3-}) \Psi' ,
\label{separclu}
\end{equation}
which proves cluster separability for $P_-$.

In this connection let us note that the semi-relativistic three-body
Hamiltonian
\begin{equation}
H = T(\vec{\rm\bf p}_1,\vec{\rm\bf p}_2,\vec{\rm\bf p}_3) + V_{123}
\ (\vec{\rm\bf p}_1,\vec{\rm\bf p}_2,\vec{\rm\bf p}_3) \, ,
\label{relat}
\end{equation}
where $T$ is the relativistic kinetic term,
\begin{equation}
T(\vec{\rm\bf p}_1,\vec{\rm\bf p}_2,\vec{\rm\bf p}_3) = \sum_{i=1}^3
\ \displaystyle{\left[ \vec{\rm\bf p}_i^2+m_i^2 \right]^{\frac{1}{2}}}
\nonumber
\end{equation}
and the potential is assumed to be the sum of two-body interactions, i. e.,
\begin{equation}
V_{123} = \sum_{i<j} V_{ij},
\nonumber
\end{equation}
does not satisfy the cluster separability condition. Indeed, if one particle is
taken far away, the Hamiltonian in the centre-of-mass system should have the
form
\begin{equation}
H^{c.m.}_{12+3} = \sqrt{\vec{\rm\bf Q}_{12}^2+\hat M_{12}^2}+
\sqrt{\vec{\rm\bf Q}_{12}^2+m_3^2}, \label{clu}
\end{equation}
where $\vec{\rm\bf Q}_{12}$ is the total momentum of the pair $(1-2)$.

On the other hand, in ref. \cite{ref15} the three-body  mass operator has been
assumed to be of the form
\begin{equation}
H^{c.m.}_{123}  = M_0+ \sum_{i<j}(\hat E_{ij} - E_{ij})
\label{separ}
\end{equation}
where $M_0$ is the overall free mass operator,
\begin{equation}
E_{ij} = \sqrt{\vec{\rm\bf Q}_{ij}^2+ M_{ij}^2} \ ~ \ ~ \ \hat E_{ij} =
\sqrt{\vec{\rm\bf Q}_{ij}^2+ \hat M_{ij}^2},
\label{separ1}
\end{equation}

and $\hat M_{ij}^2$, $M_{ij}^2$ are defined analogously to
eqs. (\ref{ciservedopo}) and (\ref{dopoancora}). Evidently the mass operator
(\ref{separ}) satisfies the cluster
separability condition (\ref{separclu}). However, if we define the mass
operator squared for the three-body system as
\begin{equation}
\hat M^2_{123} = (H^{c.m.}_{123})^2 ,
\end{equation}
this is different from
eq.~(\ref{tre-dodici}), derived from the BS equation.

It is worth noting that the mass operator (\ref{separ}), which was
considered in \cite{ref15}, commutes with the angular momentum operators
taken in the same representation as for free particles. This is not true
for the mass operator defined by eq.~(\ref{tre-dodici}). However we do not
worry about this; indeed, when considering the baryon Regge trajectories in
Sect.~5, we shall
propose a different kernel, which correponds to a
three-force with string junction and commutes with the free angular
momentum operators.

We note also that eq.~(\ref{tre-dodici}) has been considered in
ref.\cite{ref18}, however the form of this equation has been postulated, no
argument has been given for deriving it.

\section{Meson Regge trajectories}
Now we apply two- and three-body equations to the description of meson and
baryon Regge trajectories. In particular in this section we consider mesons.
We relate the $q \overline q$ interaction term $W_{12}$ (eq.
(\ref{ciservedopo})) in the two-body equation to
the potential used by Godfrey and Isgur \cite{ref3}, i. e.,
\begin{equation}
V_{12} = V^{conf}_{12}+V^{hyp}_{12}+V^{so}_{12}
\end{equation}
where
\begin{eqnarray}
V^{conf}_{12} = \displaystyle{\left[ c + \nu r - \frac{4}{3}
\ \frac{\alpha_s(r)}{r} \right] }
\nonumber\\
c = -0.253 \ ~ \  GeV, \ ~ \ ~ \ \nu = 0.18 \ ~ \ GeV^2,
\end{eqnarray}
$r$ is the distance between $q$ and $\overline q$ in the two-body cms
and $V^{hyp}_{12}$ and $V^{so}_{12}$ are, respectively, the hyperfine and
spin-orbit interactions\cite{ref3}. Squaring the mass operator
\begin{equation}
\hat M_{12} = M_{12}+V_{12},
\end{equation}
we recover eq. (\ref{ciservedopo}), with
\begin{equation}
W_{12} = M_{12}V_{12} + V_{12}M_{12} + V_{12}^2.
\label{4.int}
\end{equation}
Obviously in this case eq. (\ref{ciservedopo}) provides the same spectrum as
Godfrey and Isgur\cite{ref3}.

Here we concentrate our attention on Regge trajectories for systems of
light quarks. For large quark-antiquark separations the hyperfine and
spin-orbit
interactions (except for the Thomas precession term) can be neglected and the
main contribution comes from the confining term. On the other hand, we know
from experiment that Regge trajectories are to a high extent linear, with the
same slope. For large orbital excitations a linear
$l$-dependence (where $l$ is the orbital angular momentum) of squared masses,
i. e.,
\begin{equation}
M^2 = 8 \nu l
\label{isg-go}
\end{equation}
can be found by using a linear confining potential
in the mass operator.
Now we pose the question how to reproduce phenomenologically such a linearity
in the framework of the two-body equation (\ref{due-dodici}). If we assume
$W_{12}$ to be of the form
\begin{equation}
W_{12} (r) = \omega^2 r^{\zeta},
\label{quattro-uno}
\end{equation}
then for massless constituents and for orbital angular momenta $ l >> 1$
we may consider the semiclassical limit for $M^2$, i. e.,
\begin{equation}
M^2 = 4 { l^2 \over r^2 } + \omega^2 r^{\zeta} \ .
\label{4.2} \end{equation}
The mass squared of the system has a minimum when $r$
satisfies the equation
\begin{equation} {{dM^2} \over {dr}} = 0 , \label{4.3}
\end{equation}
whose solution is
\begin{equation}
r = ({{8 l^2} \over {\zeta \omega}^2})^{1 \over {\zeta+2}} .
\nonumber\end{equation}
In this case we have the following Regge trajectory:
\begin{equation} M^2 = 4 l^2 ({{\zeta \omega^2} \over {8 l^2
}})^{{2 \over
{\zeta+2}}} +
\omega^2 ({{8 l^2 } \over {\zeta \omega}^2})^{{\zeta \over
{\zeta+2}}} ,
\nonumber\end{equation}
which turns out to be linear when $ \zeta = 2$. In this case
\begin{equation}
M^2 = 4 \omega l . \label{4.4}
\end{equation}
Comparing eq. (\ref{4.4}) with eq. (\ref{isg-go}), we find $\omega = 2 \nu$.

This means that for large $q-\overline q$ separations the kernel $W_{12}$ is of
the oscillator form. Taking into account that all the Regge trajectories are
approximately linear, with the same slope, even for
low $l$, (see, for example,
\cite{refLiRo,ref20}), we can set the kernel (\ref{4.int})
in the form
\begin{equation}
W_{12} = W_0 + 4 \nu^2 r^2 + \Delta W_{12} (r)\,,
\label{mammamia}
\end{equation}
where $\Delta W_{12} (r)$ - which generates non-linear corrections to Regge
trajectories, e. g., spin dependent terms - is expected to be small.

We apply eq.~(\ref{due-dodici}) with the kernel (\ref{mammamia}) to
the description of $q\overline q$ Regge trajectories, where $q = u,d$ or
$s$. As it follows from the calculations of $q\overline q$ meson spectrum
done by Godfrey and Isgur\cite{ref3}, spin-dependent corrections are quite
important for $1s$ and $2s$-states. For $1p$-states they should also be
taken into account: the splitting between $1^3P_2$ and $1^3P_0$ levels is
about $200$~MeV. However for all other states those corrections are small
and can be neglected (see also the discussion at the end of this section).
If we neglect $\Delta W_{12} (r)$, the spectra of mesons composed of $u$ and
$d$
quarks are given by
\begin{equation}
M^2_{nl} = 4 m^2 + W_0 + 4\omega [2n+l-1/2] \ ~~ \ (n\geq 1) \,.
\label{meson-sp}
\end{equation}
We take the same masses of $u$ and $d$ (constituent) quarks as in
ref.~\cite{ref3}, i.e. ,
\begin{equation}
m = m_u = m_d = .22~{\rm GeV}.
\label{cane1}
\end{equation}
The two parameters $W_0$ and $\omega$ have been fixed fitting the
$\rho$-trajectory (see Fig.~1(a)), i.e.,
\begin{equation}
W_0 = -1.255 \ ~ \ {\rm GeV}^2,\qquad\quad \omega = 0.2836 \ ~ \ {\rm GeV}^2.
\label{cane2}
\end{equation}
There are eight meson Regge trajectories made of $u$ and $d$ quarks.
They are degenerate in isospin. For each isospin $I$ there are three
trajectories with angular momenta $j=l+1$, $j=l$ and $j=l-1$ and total
quark-antiquark spin
$s_{12} = 1$ and one trajectory with $j=l$ and $s_{12}=0$. We use the
notation
\begin{equation}
\nonumber
M^I_{l=j-1,s_{12} = 1},\quad
M^I_{l=j,s_{12} = 1},\quad
M^I_{l=j+1,s_{12} = 1},\quad
M^I_{l=j,s_{12} = 0}\,.
\end{equation}
All such trajectories are plotted in Figs.~1. Solid lines are determined
using eq.~(\ref{meson-sp}). We observe that all available experimental data
are in good agreement with the spectrum given by eq.~(\ref{meson-sp}) for
$l \geq 1$.
Approximate linearity with equal slopes of all nonstrange meson Regge
trajectories for $l \ge 1$ confirms $a$ $posteriori$ that nonlinear, spin
dependent terms of the potential fall down rapidly at increasing $l$. At $l =
0$ these contributions have to be taken into account, as appears in the plots.
Such deviations are smallest in the $\rho$-trajectory, suggesting that in this
case nonlinear effects are less important.

When the masses of the quark and antiquark are different, we represent the
mass operator squared in the form
\begin{equation}
\hat M^2_{12} = 4(\alpha \vecq^2 + \mu^2 ) + W_{12} +
\ G(\alpha,\mu^2,\vecq^2) \, , \label{eq-mmm}
\end{equation}
where
\begin{equation}
\mu = \Frac{m_1+m_2}{2}
\end{equation}
and
\begin{equation}
G(\alpha,\mu^2,\vecq^2) = (\sqrt{\vecq^2+m_1^2} + \sqrt{\vecq^2+m_2^2})^2
- 4(\alpha \vecq^2 + \mu^2 )\,.
\end{equation}
Note that $G$ is equal to $0$ when
\begin{itemize}
\item $m_1 = m_2$ and $\alpha = 1$ \, ;
\item $\vecq\to 0$, for any value of $m_1$ and $m_2$ \, ;
\item $\vecq^2>>m_1^2,m_2^2$ and $\alpha = 1$ \, .
\end{itemize}
If $m_1\neq m_2$, the eigenvalues and eigenfunctions of the operator
(\ref{eq-mmm}) can be suitably approximated by solving the equation
\begin{equation}
\left[ 4\left(\alpha\vecq^2+\mu^2\right) + W_{12} \right]\Psi = M^2 \Psi \, ,
\label{psi-eq}
\end{equation}
where $W_{12}$ is given by eq. (\ref{mammamia}), with $\Delta W_{12}$ = $0$. We
choose
$\alpha$ in such a way that $G$ corrections to eigenvalues of (\ref{psi-eq})
vanish to first order. This can be realized to a good approximation by imposing
$\alpha$ to be a solution of the equation
\begin{equation}
G(\alpha,\mu^2,\overline{\vecq^2}(\alpha)) = 0 \, ,
\label{alpha-eq}
\end{equation}
where $\overline{\vecq^2}(\alpha)$
is defined as the average value of $\vecq^2$ over the wave function $\Psi$,
which obviously depends itself from $\alpha$. As can be checked, eq.
(\ref{alpha-eq}) is an algebraic equation of the third degree in
$\sqrt{\alpha}$; among the roots of this equation we pick up the one which
tends to 1 as $m_1-m_2$ tends to 0. Let us examine the solution in detail. The
eigenfunctions of eq.~(\ref{psi-eq}) are
\begin{eqnarray}
\Psi_{nlm}(\vec{\bf r}) &=& \phi_{nl}(r)Y_l^m(\theta,\phi) ,
\label{aaaa1}
\\
\phi_{nl} &=& C_{nl} e^{\displaystyle -1/2\beta^2 r^2} (\beta r)^l
F\left(1-n,l+\frac{3}{2},\beta^2 r^2\right),
\label{aaaa2}
\\
C_{nl} &=& \Frac{\beta^{l+3/2}}{\Gamma(l+3/2)}
\sqrt{\Frac{2\Gamma(n+l+1/2)}{\Gamma(n)}},
\label{aaaa3}
\\
\beta^2 &=& \Frac{\omega}{2\sqrt{\alpha}}.
\label{aaaa4}
\end{eqnarray}
$\overline{\vecq^2}(\alpha)$ depends on $n,l$:
\begin{equation}
\overline{\vecq^2}(\alpha) = <nlm|{\vecq^2}|nlm>
= \left(n+l/2-1/4\right)\frac{\omega}{\sqrt{\alpha}}\, ,
\end{equation}
having denoted by $|nlm>$ the eigenstates of eq. (\ref{psi-eq}).
Therefore the solution of eq.~(\ref{alpha-eq}) which we have picked up also
depends on $n,l$ through the combination $g_{nl} = 2n+l-1/2$.
Then the eigenvalues of equation (\ref{eq-mmm}) can be approximated by
the following expression:
\begin{equation}
M^2_{nl} = 4\mu^2 + W_0 +
4\left(2n+l-1/2\right)\omega\sqrt{\alpha_{nl}} \, .
\label{eigen}
\end{equation}
The plot of $\alpha_{nl}$ as a function of $g_{nl}$ is
shown in Fig.~2 for $m_1=.22$~GeV and $m_2$ ranging from $.22$ to
$1.22$~GeV. When $m_2 \le .7$~GeV, the difference between $\alpha_{nl}$
and 1 does not exceed 10\%. In this case the correction to the slope of
the Regge trajectory does not exceed 5\% and the deviations from linearity
are negligible.

We apply (\ref{eigen}) to the description of the strange meson
trajectories. For the mass of the strange quark we used the same value as
in ref.~\cite{ref3}, i.e. $m_s = .419$~GeV. Then we can predict all the
strange meson trajectories ($M^{1/2}_{l,s_{12}}$) and mesons with hidden
strangeness ($M^0_{l,s_{12}}$) - see Figs.~3 and 4.

As in the case of non-strange mesons, the agreement between our theoretical
predictions and the existing experimental data\cite{ref21} is good,
except for the states with $l=0$.

The spectra given by eq.~(\ref{eigen}) contain also radial excitations. The
mass of the $q\overline q$ state with $n=2,~l=0$ is predicted around
$1.7$~GeV, while the in pseudoscalar channel the first radial excitation is
around $1.3$~GeV ($\pi(1300)$ and $\eta(1295)$) and in the vector channel it is
around $1.4$~GeV ($\omega(1390)$ and $\rho(1450)$). This discrepancy can be
due to neglecting, in our model, spin-spin interaction, which mixes states
with different $n$.

\section{Baryon Regge trajectories}

Let us apply now the three-body null plane equation (\ref{tre-dodici}) to
the description of the baryon Regge trajectories.
We consider a system of three light  quarks
in a colour singlet state.
For orbital excitations with large angular momenta we may distinguish among
three different configurations:

\begin{enumerate}
\item String-like configuration, with two quarks forming a
diquark ( $r_{12} \sim r_{23} \gg r_{31}$, where $r_{ij}$ is the
relative distance between the $i-$th and the $j-$th quark in the
two-particle cms) [``D''-configuration];
\item Symmetric triangle-like  configuration ( $r_{12} \sim r_{23}
\sim r_{31}$ ) [``T''-configuration];
\item Star-like configuration with string junction
[``Y''-configuration].
\end{enumerate}

In the first case, the problem reduces to the two-body case; in
particular $\omega$ is the same as in quark-antiquark system,
since in a baryon the diquark is in a colour anti-triplet state. For
rather large overall angular momenta, when the rest mass of the diquark may be
neglected, the Regge trajectory is the same as in two-body case,
eq.~(\ref{4.4}).

In the second case we recover eq.~(\ref{tre-dodici}), where the
two-body kernels $W_{ij}$ are of the oscillator form. Taking
into account the colour degree of freedom yields
\begin{equation}
W_{ij} = {1 \over 2} \omega^2 r_{ij}^2 .
\nonumber
\end{equation}

If, moreover, we take into account of the symmetry of the
configuration and proceed similarly to the two body case, we
find
\begin{equation} M^2 = 6 \omega L_{123}\,, \label{4.6}
\end{equation}
where $L_{123}$ is the overall angular momentum of the three-quark system.
Therefore for a fixed $L_{123}$ the diquark-quark configuration has a lower
energy than the symmetric triangle configuration, according to the following
ratio:
\begin{equation}
M^2_T : M^2_D = 1.5 : 1\,.
\nonumber\end{equation}

In the case of a relativized three-body Schr\"odinger equation, i. e.,
\begin{equation}
(\sum_{i=1}^3 \sqrt{\vec{\rm\bf p}^2_i+m_i^2} + V_{123}) \Psi' =  M \Psi' ,
\end{equation}
where $V_{123}$ is linear either with respect
to relative distances between quarks (triangle configuration) or
with respect to distances of quarks from the string centre (string junction
configuration), the situation is different; in this case it results\cite{ref22}
\begin{equation}
M^2_T : M^2_D  =  1.3 : 1.0 .
\nonumber\end{equation}
The difference in the ratio $M^2_T : M^2_D$ is due to the
relativistic recoil effects taken into account in the three-body null plane
equation.

In connection with triangle configuration we recall, as shown in sect. 3,
that the three-body semi-relativistic Hamiltonian with a two-body potential
does not satisfy cluster separability condition. Alternatively, using
three-body mass operator (\ref{separ}) - which satisfies cluster separability
condition - would lead to
a non-linear Regge trajectory. In fact, if we assume the potential $V_{ij}$,
which
describes quark-quark interaction in the two-quark cms, to be proportional to
the distance between the two interacting particles - as imposed by
linearity of meson Regge trajectories -, the same potential in the overall cms
is no longer linear, since, according to eqs. (\ref{separ}) and (\ref{separ1}),
it transforms to
\begin{equation}
V_{ij}' = \hat E_{ij} - E_{ij}
\end{equation}
where
\begin{equation}
\hat E_{ij} = \sqrt{\vec{\rm\bf Q}_{ij}^2+(M_{ij}+V_{ij})^2} \, , \ ~~~ \
E_{ij} = \sqrt{\vec{\rm\bf Q}_{ij}^2+ M_{ij}^2}
\end{equation}
and $M_{ij}$ is the two-body free mass operator, defined analogously to
(\ref{dopoancora}).

In the case of a star-like configuration with string junction
the interaction is not a two body one. This is a genuine three-body
force. Indeed arguments in favour of a three-body force for describing the
baryon spectrum have been illustrated in ref.~\cite{ref23}. Also in the
framework of a relativized Schr\"odinger equation a three-body force with
string
junction ~\cite{refCaIs2} allows a unified treatment of mesons and baryons,
whereas exclusive
use of two-body forces ~\cite{ref5} demands a quark-diquark structure.
Then we assume a kernel of the form
\begin{equation}
U_{123} = U_0 + \omega'^2 (r_{01}^2+r_{02}^2+r_{03}^2) ,  \ ~~ \
r_{0i}=|\vec{\rm\bf r}_{0i}|, \ ~~~ \ (i=1,2,3) ,
\label{3b-pot}
\nonumber\end{equation}
where $U_0$ is a constant and
$\vec{\rm\bf r}_{0i}$ is the vector from the string center to the $i-$th
quark in the overall cms. It is not difficult to recognize that such
a kernel is of the type ~(\ref{tre-quattro}), since $\vec{\rm\bf r}_{0i}$ is
conjugate to momentum
\begin{equation}
\vec{\rm\bf p}_{0i} = \vec{\rm\bf p}_i-\vec{\rm\bf p}_0,
\nonumber\end{equation}
where
\begin{equation}
\vec{\rm\bf p}_l \equiv ({\rm\bf p}_{lT}, p_{lz}) \ ~~~ \ (l = 0,1,2,3)
\end{equation}
and the $p_{lz}$ are related to the fractional momenta $\xi_l$ by means of
the following equations:
\begin{equation}
\xi_l = {{\sqrt{m^2_l+\vec{\rm\bf p}_l^2} + p_{lz}} \over M},
\ ~~ \ (l = 0,1,2,3).
\end{equation}
The kernel (\ref{3b-pot}) contains also the dependence on
c.m. coordinate. To eliminate it, we proceed as in the non relativistic
case of three particles with equal masses, defining the relative
coordinates as follows:
\begin{eqnarray}
\vec R  &=& \Frac{1}{3}\left(\vec{r_{01}} + \vec{r_{02}} +
\vec{r_{03}}\right)\nonumber\\
\vec\rho &=& \vec{r_{01}} - \vec{r_{02}}\\
\vec\lambda &=& \Frac{1}{3}\left(\vec{r_{01}} + \vec{r_{02}} -
2\vec{r_{03}}\right)\nonumber\,.
\end{eqnarray}
In order to take into account relativistic effects, we write the interaction
term in the form
\begin{equation}
U_{123} = U_0 + {\omega^\prime}^{\,2}\left(\Frac{2}{3}\eta_\lambda
{\vec\lambda}^2
+\Frac{1}{2}\eta_{\rho} {\vec\rho}^2\right) \, .
\label{5-11}
\end{equation}
In the non relativistic limit the factors $\eta_\lambda$ and
$\eta_\rho$ tend to 1. Let us stress a difference between our model
and the non-relativistic harmonic oscillator model. In our case the
interaction term (\ref{5-11}) appears in the mass operator squared, i. e.,
\begin{equation}
\hat M_{123}^2 = \left[\sqrt{\vecqq^2 + M^2_{12}(\vecq^2)} +
\sqrt{\vecqq^2 + m_3^2} \ ~ \right]^2 + U_{123},
\label{5-12}
\end{equation}
(we have used shortened notation $\vecqq$ for $\vecqq_{12}$),
while in the non relativistic case it is a part of the mass operator.
Therefore the mass spectrum and the Regge trajectories in the relativistic
and non relativistic models with harmonic oscillator interaction will be
essentially different for systems with light quarks. The momenta $\vecqq$
and $\vecq$ in equation (\ref{5-12}) are, respectively, the momenta of
the subsystems $(12)+3$ and $1+2$ in their c.m. frames.

In the nonrelativistic case $\vec\lambda$ and $\vec\rho$ are the conjugate
coordinates, respectively, to the momenta $\vecqq$ and $\vecq$.
In the relativistic case this can be justified
only for $\vec\lambda$. Of course the value of $|\vec\lambda|$ is not
Lorentz invariant and in the overall c.m. system it can be regarded as
the distance between the subsystem $(1+2)$ and particle $3$.
In line with these considerations we can take the coefficient
$\eta_\lambda$ in eq.~(\ref{5-11}) equal to 1. The parameter $\eta_\rho$,
which describes the relativistic recoil effect for subsystem $1+2$, will be
fixed later. Therefore in the relativistic case the choice of the
kernel (\ref{5-12}), where both $\vec\lambda$ and $\vec\rho$
are considered as conjugate coordinates to the momenta $\vecqq$ and $\vecq$
should be treated as an "ansatz", which however does not correspond to a sum
of two-body interactions. An important consequence of this choice of the
interaction is that it commutes with the angular momentum operators taken
in the same representation as for free particles:
\begin{equation}
\vec L_{123} = \vec L  + \vec l  =
-i\vecqq\times\Frac{\partial}{\partial\vecqq}
-i\vecq\times\Frac{\partial}{\partial\vecq}.
\label{5-11'}
\end{equation}
Since the mass and angular momentum operators have been fixed according to eqs.
(\ref{5-11}) and (\ref{5-11'}) respectively, all other Poincar\'e generators
can
be constructed according to the scheme prposed by Berestetsky and
Terent'ev\cite{refBeTe} (see also \cite{ref15}). The angular condition is
satisfied, therefore the eigenstates of our Hamiltonian are guaranteed to be
the null plane restrictions of covariant wavefunctions\cite{ref6}.

Let us consider the spectrum predicted by the mass operator squared
(\ref{5-12}). Using the results of the previous section, we write the
kinetic term in eq.~(\ref{5-12}) as follows:
\begin{eqnarray}
\left[\sqrt{\vecqq^2 + M^2_{12}(\vecq^2)} +
\sqrt{\vecqq^2 + m_3^2} \ ~ \right]^2 &=&
4(\alpha \vecqq^2 + \mu^2) + G(\alpha,\mu^2,\vecqq^2),\\
\mu &=& \Frac{1}{2}\left[M_{12}(\vecq)+m_3\right],
\end{eqnarray}
and we assume $m_1=m_2=m$. As in the two-body case, we approximate the
eigenvalues
and eigenfunctions of the operator $\hat M^2_{123}$( eq. (\ref{5-12}) ) with
those of the eigenvalue equation
\begin{equation}
\left[
4(\alpha \vecqq^2 + \mu^2) +
{\omega^\prime}^{\,2}\left(\Frac{2}{3}{\vec\lambda}^2
+\Frac{1}{2}\eta_{\rho} {\vec\rho}^2\right) + U_0
\right]\Psi = M^2\Psi\,,
\label{3eige}
\end{equation}
where $\alpha$ solves the equation
\begin{equation}
G(\alpha,\mu^2(\overline{\vecq^2}),\overline{\vecqq^2}) = 0\,.
\end{equation}

Here $\overline{\vecq^2}$ and $\overline{\vecqq^2}$ are the average values of
$\vecq^2$ and $\vecqq^2$ for the eigenfunctions of eq.~(\ref{3eige}), which
can be written in the form
\begin{equation}
\Psi_{NLM,nlm}\left(\vecqq,\vecq\right) =
\psi_{NLM}\left(\vecqq\right)
\phi_{nlm}\left(\vecq\right)\,, \ ~~ \ (N,n \geq 1)
\label{solution}
\end{equation}
where the functions $\psi_{NLM}\left(\vecqq\right)$ and
$\phi_{nlm}\left(\vecq\right)$ are defined by the expressions
(\ref{aaaa1}-\ref{aaaa4}), with
\begin{equation}
\beta_Q^2 = \Frac{1}{\sqrt{6}}\Frac{\omega^\prime}{\sqrt{\alpha}}\,, \ ~ \
\ ~ \beta_q^2 = \Frac{\omega^\prime}{2\sqrt{2}}\sqrt{\eta_\rho}.
\end{equation}
Assuming $\overline{M_{12}^2(\vecq)}>>m_3^2$, and moreover that $U_0 = 2W_0$,
where $W_0$ is the same as for $q\overline q$ system (which amounts to taking
$W_0$ per degree of freedom),  we can write the eigenvalues of
eq.~(\ref{3eige}) as follows:
\begin{equation}
M_{NLnl}^2 = M^2_{nl} + W_0 + 4\omega^\prime
\sqrt{\Frac{2}{3}\alpha_{NL}}\left(
2N+L-1/2\right)\,,\label{spectra}
\end{equation}
where
\begin{equation}
M^2_{nl} = W_0 + 4m^2 + 2\sqrt{2} \omega^\prime \sqrt{\eta_\rho} (2n+l-1/2)
\,.  \label{dimass}
\end{equation}
Let us consider the limiting configuration of $L\to\infty$ and
$l$
fixed, which corresponds to a string-like diquark-quark configuration. In this
case, since $\vecqq^2>>M_{12}^2,m_3^2$ and $\alpha_{NL}\to 1$, the slope of the
Regge trajectory for this configuration results to be
$\sqrt{3/2}(4\omega^\prime)^{-1}$; this slope equals the slope of the meson
trajectory, therefore
\begin{equation}
\omega^\prime = \sqrt{3/2} \omega \,.
\end{equation}
On the other hand, in the limit $l\to\infty$, with $L$
fixed, the slope of the Regge trajectory results to be
$ (2\sqrt{3}\omega\sqrt{\eta_\rho})^{-1}$.
As the two slopes are equal, $\eta_\rho$ is uniquely fixed to $4/3$.
Therefore the mass spectrum (\ref{spectra}) has the form
\begin{equation}
M^2_{NLnl} = \{ -2 W_0 + 4 m^2 + 4 \omega[\sqrt{\alpha}(2N+L)+2n+l
\ -1/2(1+\sqrt{\alpha})] \} \ ~ \ GeV^2  \,,
\label{5-24}
\end{equation}
where $W_0$, $m$ and $\omega$are the same as for $q \overline q$ sector (see
eqs (\ref{cane1}) and (\ref{cane2})).
After symmetrization over all quarks the wave-function (\ref{solution}) can
be considered a good approximation to the ground state of the
$\Delta$-isobar. According to eq.~(\ref{5-24}), the mass of the ground
state ($N=n=1$, $L = l = 0$) in this case is equal to
$M_{1010} = 1.043$~GeV,
which is intermediate between the masses of the nucleon and of the
$\Delta$-isobar. The spin-spin interaction will increase the mass for a total
angular momentum $J=3/2$ and decrease it for $J=1/2$.

In the description of Regge trajectories $M_{nl}$ can be considered as the
effective mass of a diquark. According to eq.~(\ref{dimass}) for $n=1$ and
$l=0$ we have $M_{10} = 0.8$~GeV. This value can be considered as an
estimation of the mass of a diquark in the ground state at $n=1$ and $l=0$.
To take into account spin-spin
interaction we shall fit the baryon Regge trajectories considering the mass
of the diquark as a free parameter. As we have already seen for mesons, the
spin-spin interaction is particularly
important in the $s$-wave. Therefore, if we consider the states with $l=0$
and $L \ge 1$, the spin-spin interaction between the third quark and the
two others may be neglected.

Let us classify the baryon Regge trajectories composed of $u,d$ quarks in
the diquark-quark picture. The lowest states are composed of $s$-wave
diquarks: ${\cal D}_{00}(I_{12}= s_{12} = 0)$ and
${\cal D}_{11}(I_{12}= s_{12} = 1)$. There are two nucleon Regge
trajectories corresponding to the orbital excitations of the $q-{\cal
D}_{00}$ system with $I=1/2$: we denote them by $N_{L=J-1/2}$ and
$N_{L=J+1/2}$. Six Regge trajectories correspond to the orbital excitation
of the $q-{\cal D}_{11}$ system. They are degenerated in the isospin
$I=1/2,3/2$: $(N-\Delta)_{L=J-3/2}$, $(N-\Delta)_{L=J+1/2}$,
$(N-\Delta)^\prime_{L=J+1/2}$ and $(N-\Delta)_{L=J+3/2}$. The trajectories
$(N-\Delta)$ and $(N-\Delta)^\prime$ correspond to different total spins of
quarks, $S_{123} = 3/2$ or $1/2$. To describe all these trajectories, we use
the two body equation (\ref{eigen}), where the mass of the quark was
taken as for meson trajectories, i.e., $m_u=m_d=.22$~GeV and the masses of
${\cal D}_{00}$  and ${\cal D}_{11}$ diquarks have been found from fits to
the main $N$ and $\Delta$ trajectories, $N_{L=J-1/2}$ and
$\Delta_{L=J-3/2}$ (see Figs.~ 5 and 6), yielding $M_{00} =.44$ GeV and $M_{11}
=.80$ GeV, which are consistent with the values given above.

The agreement of the theoretical predictions with the available
experimental data is quite satisfactory. We are able also to predict the
$\Lambda$ Regge trajectory composed of $s-{\cal D}_{00}$ orbital
excitations. Our predictions (see Fig.~6) are in very good agreement with
data.

\section{Conclusions}

Starting from the BS equation in ZRA, we have introduced a null
plane wave-function and have derived a three-dimensional two-
body equation. We have generalized
this equation to the case of finite range interactions and have discussed
the choice of angular momentum operators in this scheme. We have
considered also the three-body problem in the same approach, showing
that in ZRA it is possible to derive a three-body null
plane Hamiltonian with two-body interaction
which satisfies cluster separability condition. We have generalized also this
approach to the case of two- and three-body interactions with finite
range. As an application of this scheme we have considered Regge
trajectories for relativistic two- and three-quark systems. We
have shown that linear trajectories can be derived when the
interaction term in the mass square operator is of the oscillator
form.

We have discussed the relation between this kernel and the
Hamiltonian used by Godfrey and Isgur\cite{ref3}. Then we have applied our
model
to the description of the meson Regge trajectories $\bar q - q$, $\bar q -
s$ ($\bar s - q$) and $\bar s - s$. The model describes, with only four free
parameters, $m_u = m_d = m$, $m_s$, $W_0$ and $\omega$, all orbitally excited
meson states in a quite satisfactory way. We have considered
also the choice of the kernel in the three-body case. It
appear that the choice of two-body forces in the relativistic three-body
equation satisfying the cluster separability property leads to an equation
which is essentially different from the semi-relativistic generalizations of
the Schr\"odinger equation.

We have also considered a kernel which corresponds to a
three-body force with string junction; this operator can be taken in a form
which commutes with the angular momentum operator in the free particle
representation, so that the angular condition is satisfied.
Choosing this model of interaction, we consider
Regge trajectories for three-quark systems. We calculate all baryon Regge
trajectories composed of $u$ and $d$ quarks in the diquark-quark configuration
introducing phenomenologically hyperfine splitting through the difference in
masses of the two color antitriplet diquarks, ${\cal D}_{00}$  and ${\cal
D}_{11}$. Our predictions are in good agreement with $N$ and $\Delta$ Regge
trajectories. Also the prediction of the $\Lambda$ Regge trajectory, which
corresponds to $s-{\cal D}_{00}$ orbital excitations, is in very good
agreement with data.

To summarize, the approximation proposed for the BS equation or for the
four-dimensional three-body equation turns out to be a covariant Hamiltonian
dynamics\cite{ref6}, provided we consider a convenient representation of the
angular momentum operators and choose a kernel which commutes with such
operators. Moreover, concerning quark dynamics, this approximation
may be implemented by three-body interactions with string junction - whose
importance has been already stressed in the literature - to yield a unified
description of meson and baryon Regge trajectories.

\centerline{ACKNOWLEDGMENTS}

{}~~~~~~~~~~~~~~~~~~

The authors are very grateful to their friends M.M. Giannini, A.B. Kaidalov and
Yu.A. Simonov for numerous stimulating discussions.

\hbox {FIGURE CAPTIONS \hfill}
\noindent

[Fig. 1] - Meson Regge trajectories made of $u$- and/or $d$-quarks
($M^I_{L,s_12}$). \newline

[Fig. 2] - The parameter $\alpha_{nl}$ as a function of $g_{nl} = 2(n-1)+l$,
found from the solution of eq. (\ref{alpha-eq}) for $m_1$ = $.22$ GeV and
$m_2$ ranging from $0.22$ to $1.22$ GeV. \newline

[Fig. 3] - Regge trajectories for strange mesons ($M^{1/2}_{L,s_12}$). \newline

[Fig. 4] - Regge trajectories for $\overline s s$ mesons
($M^0_{L,s_12}$). \newline

[Fig. 5] - Non-strange baryon Regge trajectories. \newline

[Fig. 6] - $\Lambda$ Regge trajectory, corresponding to the orbital
excitations of the $s- \cal{D}_{00}$ system. \newline


\newpage

\begin{thebibliography}{99}



\bibitem {ref1} T.Appelquist and H.D.Politzer, Phys.Rev.Lett. : \em
34\rm , 43 (1975); Phys.Rev.:\em  D 12\rm ,1404 (1975);
E.Eichten et al.,Phys.Rev.Lett.:\em  34\rm , 369 (1975);
T. Appelquist et.al., Phys.Rev.Lett.:\em 34 \rm , 365 (1975).

\bibitem {ref2} S.N.Mukherjee et. al.: Phys.Rep. \em  231\rm , 201 (1993)

\bibitem {ref3} S.Godfrey and N.Isgur: Phys.Rev. \em  32D\rm , 189 (1985)

\bibitem {refCaIs2} S.Capstick and N.Isgur: Phys.Rev. \em  34D\rm , 2809 (1986)

\bibitem {ref4} J.L.Basdevant and S.Bukraa: Z.Phys. \em C28\rm , 413 (1985)

\bibitem {ref5} J.L.Basdevant and S.Bukraa: Z.Phys. \em C30\rm , 103 (1986)

\bibitem {ref6} H.Leutwyler and J.Stern: Ann.Phys. \em  112\rm
, 490 (1979)

\bibitem {ref7} B.D.Keister and W.N.Polyzou: Adv.Nucl.Phys.\em 20\rm,
225, Ed. J.W.Negele and E.Vogt (1991)

\bibitem {ref8} L.A.Kondratyuk: Few-Body Systems, Suppl.\em  6\rm , 512 (1992)

\bibitem {ref9} K.Hornbostel, S.J.Brodsky, and H.-C.Pauli : Phys.Rev.
\em  D 41\rm , 3814 (1990); N.A.Aboud and John R. Hiller: Phys.Rev.
\em  D 41\rm , 937 (1990); St.Glazek and M.Sawicki : Phys.Rev.
\em  D 41\rm , 2563 (1990); J.R.Hiller : Phys.Rev. \em  D 44\rm , 2504 (1991);
St.Glazek and C.M.Shakin : Phys.Rev. \em  C 44\rm , 1012 (1991);
C.M.Yung and C.J.Hamer : Phys.Rev. \em  D 44\rm , 2598 (1991);
A.Harindranath and Robert J. Perry : Phys.Rev. \em  D 43\rm ,492 (1991);
D.Mustaki, S.Pinsky, J.Shigemitsu, and K.Wilson : Phys.Rev.
\em  D 43\rm , 3411 (1991); R.J.Perry and A.Harindranath: Phys.Rev.
\em  D 43\rm , 4051 (1991)

\bibitem{ref9a} H.H.Liu and D.E.Soper: Phys.Rev.
\em  D 48\rm , 287 (1993)

\bibitem {ref10} S.Weinberg: Phys.Rev. \em  150\rm , 1313
(1966)

\bibitem {ref11} M.V.Terent'ev: Yad.Fiz. \em  24\rm , 207 (1976)

\bibitem {refKoTe} L.A.Kondratyuk and M.V.Terent'ev: Yad.Fiz. \em  31\rm ,
\ 1087 (1980)

\bibitem {ref12} G. t'Hooft: Nucl.Phys. \em  B 75\rm , 461 (1974)

\bibitem {ref13} V.A.Karmanov: Sov.Journ.Nucl.Part.Phys. \em  B 19\rm ,
228 (1988); V.A.Karmanov and A.V.Smirnov: Nucl.Phys. \em A 546\rm , 691 (1992)

\bibitem {ref14} M.M.Giannini, L.Kondratyuk and P.Saracco: to appear on
Few Body System


\bibitem {ref15} B.L.G.Bakker, L.A.Kondratyuk and M.V.Terent'ev:
Nucl.Phys.
\em  B 158\rm , 497 (1979)

\bibitem {refCaIs} S.Capstick, S.Godfrey, N.Isgur and J.Paton: Phys.Lett.
\em  B 175\rm , 457 (1986)

\bibitem {refBeSa} H.A.Bethe and E.E.Salpeter: "Quantum Mechanics of One and
Two Electron Atoms", Springer-Verlag, Berlin, 1957

\bibitem {refItZu} C.Itzykson and J.B.Zuber: "Quantum Field
Theory", McGraw-Hill Book Co, New York, 1980

\bibitem {refFey} R.P.Feynman, M.Kislinger and F.Ravndal: Phys.Rev.
\em  D 3\rm , 2706 (1971)

\bibitem {ref17} H.J.Melosh: Phys.Rev. \em  D 9\rm , 1095 (1974)

\bibitem {ref18} L.A.Kondratyuk and L.V.Shevchenko:
Yad.Fiz. \em  29\rm , 792 (1979)

\bibitem {ref19} F.Iachello: Phys.Rev.Lett. \em 62 \rm, 2440 (1989);
Nucl.Phys. \em  A 518\rm , 173 (1990); F.Iachello, N.C.Mukhopadhyay and
L.Zhang:
Phys.Rev. \em D 44\rm , 898 (1991)

\bibitem {refLiRo} F.Lizzi and C.Rosenzweig: Phys.Rev. \em  D 31\rm , 1685
(1985)

\bibitem {ref20} A.B.Kaidalov: Sov.Journ.Nucl.Phys. \em  B 51\rm , 319 (1990)

\bibitem {ref21} Particle Data Group; K.Hikasa et al: Phys.Rev. D
\em  45\rm
(1992)

\bibitem {ref22} A.Martin: Z.Phys. C \em  32\rm , 355 (1986)

\bibitem {ref23} M.M.Giannini: NuovoCim. A \em 76\rm, 455 (1976); D.Drechsel,
M.M.Giannini and L.Tiator: Few Body Systems, Supplementum \em 2\rm , 448 (1987)

\bibitem {refBeTe} V.B.Berestetski and M.V. Terent'ev: Yad.Fiz. \em 25\rm, 653
(1977)

\end{thebibliography}
\end{document}